\documentclass[11pt,a4paper]{article}
\usepackage{fullpage}
\usepackage{amsmath, amsthm, amsfonts, amssymb, latexsym}
\usepackage{mathtools} 
\let\originalleft\left
\let\originalright\right
\renewcommand{\left}{\mathopen{}\mathclose\bgroup\originalleft}
\renewcommand{\right}{\aftergroup\egroup\originalright}

\usepackage[colorlinks, citecolor=Red, linkcolor=Blue, urlcolor=blue]{hyperref}
\usepackage{caption}
\usepackage[dvipsnames]{xcolor}
\usepackage{tikz}
\usepackage{enumerate}
\usepackage{cancel}
\usepackage{dsfont}
\usepackage{accents}
\usepackage{tensor}

\usepackage{authblk}

\newcommand{\wrt}{w.r.t.\ }
\newcommand{\cf}{cf.\ }

\makeatletter
\DeclareRobustCommand{\lambdabar}{\mathord{%
 \text{$\m@th\mkern+1mu\raisebox{-0.2ex}[0pt][0pt]{$\mathchar'26$}\mkern-10mu \lambda$}%
}}
\makeatother

\newcommand{\EV}[1]{\langle #1 \rangle}

\newcommand{\ud}{\mathrm{d}}
\newcommand{\del}{\partial}

\newcommand{\R}{\mathbb{R}}

\newcommand{\1}{\mathbbm{1}}

\newcommand{\eps}{\varepsilon}

\def\1{{\mathds{1}}}

\newcommand{\const}{{\mathrm{const}}}

\newcommand{\nn}{\nonumber}
\newcommand{\beq}{\begin{equation}}
\newcommand{\eeq}{\end{equation}}

\newcommand{\defeq}{\mathrel{\coloneq}}

\newcommand{\cO}{{\mathcal{O}}}

\begin{document}

\title{Quantum fields on self-similar spacetimes}
\author{Jochen Zahn\thanks{jochen.zahn@itp.uni-leipzig.de} \\ Institut f\"ur Theoretische Physik, Universit\"at Leipzig\\ Br\"uderstr.\ 16, 04103 Leipzig, Germany}

\date{\today}

\maketitle

\begin{abstract}
We study scalar (not necessarily conformal) quantum fields on self-similar spacetimes. It is shown that in states respecting the self-similarity the expectation value of the stress tensor gives rise to a Lyapunov exponent $\omega_q = 2$, with a leading coefficient which is state-independent and geometric. Three examples for states respecting self-similarity are presented. 
\end{abstract}

\section{Introduction}

In a gravitational collapse scenario, one generally expects the appearance self-similar structures \cite{Eardley1974}, in particular for collapse leading to the formation of naked singularities. This expectation has been confirmed in many collapse scenarios \cite{OriPiran1987, Choptuik, AbrahamsEvans1993, EvansColeman}, see also \cite{GundlachHilditchMartin-Garcia} for a recent review of the subject. Hence, self-similar spacetimes are highly relevant in the context of the weak cosmic censorship conjecture \cite{PenroseCosmicCensorship, WaldCosmicCensorship}.

Depending on the matter type, self-similarity can be either discrete (as for the minimally coupled scalar field considered in \cite{Choptuik}) or continuous (as for perfect fluid matter considered in \cite{OriPiran1987, EvansColeman}). This means that there are coordinates $(u, y^i)$ such that
\beq
\label{eq:SelfSimilarity}
 g_{\mu \nu}(u + \alpha, y^i) = e^{- 2 \alpha} g_{\mu \nu}(u, y^i),
\eeq
where for discrete self-similarity $\alpha$ is restricted to integer multiples of some period $\Delta$, while for continuous self-similarity this holds for any real $\alpha$.
In either case, for increasing $u$, distances shrink, while curvature blows up, leading to a singularity as $u \to \infty$ (unless the curvature of $g_{\mu \nu}$ happens to vanish). In such a regime, one expects quantum effects to become relevant. This motivates the study of quantum fields on self-similar spacetimes.

The basis of most studies of quantum effects in self-similar spacetimes is the semi-classical Einstein equation
\beq
\label{eq:SCEE}
 \frac{1}{8 \pi} G_{\mu \nu} = T^{\mathrm{class}}_{\mu \nu} + \langle T_{\mu \nu} \rangle,
\eeq
with $T^{\mathrm{class}}_{\mu \nu}$ the stress tensor of classical matter and $\langle T_{\mu \nu} \rangle$ the expectation value of the quantum stress tensor in some quantum state. However, computing $\langle T_{\mu \nu} \rangle$ requires renormalization, which is notoriously difficult already for stationary spacetimes which are explicitly given (but see \cite{ZilbermanEtAl2022, ArrecheaEtAl2024} for recent progress in the context of black hole spacetimes). The problem becomes even harder when the spacetime is neither stationary nor given analytically, as expected for full solutions to \eqref{eq:SCEE}. Different strategies have been used to tackle these difficulties, partly with contradictory result (see the Introduction of \cite{TomasevicWu} or Section~IV.D of \cite{GundlachHilditchMartin-Garcia} for an overview). One possibility is to work in two spacetime dimensions, where renormalization is much easier to perform \cite{StromingerThorlacius, PelegBoseParker, ChibaSiino, AshtekarPretoriusRamazanoglu, BenitezEtAl2020, TomasevicWu}. One problem with this approach is that even if the two-dimensional theory arises from dimensional reduction (exploiting spherical symmetry) of a four-dimensional theory, it is unclear whether the two-dimensional model correctly captures the features of the four-dimensional quantum field theory, due to ``dimensional reduction anomalies'' \cite{FrolovSuttonZelnikov, BalbinotEtAl2001} (see also Section~VI of \cite{ArrecheaEtAl2023} for a recent comparison of exact stress tensor expectation values with those obtained via dimensional reduction on black hole spacetimes). In fact, we will see that while the dimensional reduction technique used in \cite{TomasevicWu} gives the same exponential growth of stress tensor expectation values that we find, the coefficients of that growth are of a qualitatively different form.

A further possibility to solve \eqref{eq:SCEE} is not to compute any expectation value at all, but to make some ansatz for $\langle T_{\mu \nu} \rangle$, which is supposed to capture relevant quantum effects. This allows to solve \eqref{eq:SCEE} numerically, but the results obviously depend very much on the ansatz for $\langle T_{\mu \nu} \rangle$. For example, in \cite{AyalPiran} the ansatz for $\langle T_{\mu \nu} \rangle$ was meant to incorporate Hawking evaporation effects, but did not constitute a conserved tensor.

In recent years, also full solutions of the semi-classical Einstein equation \eqref{eq:SCEE} were attempted numerically \cite{BercziEtAl2022}. In this approach, the renormalization is performed by a Pauli-Villard scheme, i.e., additional ``ghost'' fields are introduced which cancel the divergences. In principle, one must consider the limit in which the masses of the ghost fields become infinite, but in practice, finite masses are used. This may be justified as long as the curvature  is negligible with respect to the ghost mass. But in critical collapse the curvature becomes arbitrarily large. Hence, it seems questionable whether a Pauli-Villard renormalization scheme is appropriate for the investigation of quantum effects in critical collapse. 

The approach taken in \cite{BradyOttewill} towards solving \eqref{eq:SCEE} for self-similar spacetimes is to analyse the general structure of $\langle T_{\mu \nu} \rangle$ under suitable conditions. We will follow the same general strategy but relax an assumption that was crucial for the argument used in \cite{BradyOttewill}, namely conformal coupling of the quantum field. We will see that, by using a different argument, the result of \cite{BradyOttewill} can be easily generalized to non-conformal coupling. Specifically,  we consider the massless scalar field, i.e., subject to the field equation
\beq
 \nabla^\mu \nabla_\mu \phi - \xi R \phi = 0,
\eeq
where $\xi$ parameterizes the coupling to curvature ($\xi = 0$ corresponding to minimal and $\xi = \frac{1}{6}$ to conformal coupling). We assume the existence of a state that respects self-similarity in the sense that the Wightman two-point function $w(x; x') = \langle \phi(x) \phi(x') \rangle$ fulfills
\beq
\label{eq:2ptSelfSimilar}
 w( u + \alpha, y ; u' + \alpha, y' ) = e^{2 \alpha} w(u, y; u', y').
\eeq
Motivations for the consideration of states respecting self-similarity will be given below.
We will not prove the existence of such a state for generic self-similar spacetimes, but in three concrete examples: A certain patch of Minkowski space is self-similar, and the vacuum state respects this self-similarity. We explicitly construct states respecting self-similarity on the Hayward spacetime \cite{HaywardSpacetime} and the critical Roberts spacetime \cite{Roberts1989}, which are self-similar solutions to the Einstein-scalar field equations. However, the construction performed on the critical Roberts spacetime allows for straightforward generalization to other self-similar spacetimes.

The consideration of states respecting self-similarity can be motivated as follows: In order to focus on generic and unavoidable quantum effects, one would like to choose a state that respects the self-similarity as well as possible (so as not to introduce deviations from self-similarity that are avoidable). Now the anti-symmetric part of the two-point function is given by $i \Delta(x; x')$ with $\Delta(x; x')$ the difference of retarded and advanced propagators for the Klein-Gordon operator. These are uniquely determined by the geometry and have precisely the scaling behaviour \eqref{eq:2ptSelfSimilar} under self-similarity transformations. Hence, the antisymmetric part of $w$ must fulfill \eqref{eq:2ptSelfSimilar}. But the symmetric part of $w$ can be bounded from below by the antisymmetric part through the positivity requirement \cite{KayWald91}, and thus must scale at least as strong as the antisymmetric part. Hence, our requirement \eqref{eq:2ptSelfSimilar} enforces the minimal scaling compatible with commutation relations and positivity. Furthermore, states respecting self-similarity naturally arise by mode sum constructions of the two-point function, as seen in examples below. Finally, we will argue (though not rigorously prove) that the leading behaviour of the stress tensor that we find for states respecting self-similarity is generic for a large class of other states.

Now for a self-similar spacetime, a shift in $u$ by $\alpha$ yields the same metric, but multiplied with a constant scale factor $e^{- 2 \alpha}$. But the behaviour of renormalized expectation values under such a constant scale transformation is well controlled: When the two-point function is scaled accordingly (corresponding to our requirement \eqref{eq:2ptSelfSimilar}), then renormalized expectation values scale almost homogeneously, i.e., the expectation value scales according to its canonical dimension, up to an additional term (involving the logarithm of the scale factor), which is state-independent and geometric. This behaviour of the renormalized expectation value under a change of scale seems to have first been made explicit in \cite{WaldTraceAnomaly} for the conformally coupled case, but it was implicitly already contained (for general coupling) in \cite{Christensen76} (see also \cite{HackThesis}, Theorem IV.2.1, for an explicit statement in the general case). The upshot is that, under the assumption \eqref{eq:2ptSelfSimilar} on the state, the expectation value of the stress tensor behaves under ``self-similarity time'' shift as
\beq
\label{eq:MainResult}
 \langle T_{\mu \nu} \rangle ( u + \alpha, y ) = e^{2 \alpha} \left( \langle T_{\mu \nu} \rangle(u, y) + \frac{\alpha}{4\pi} V_{\mu \nu}(u, y) \right),
\eeq
as will be shown below.
Here $V_{\mu \nu}$ is state-independent and independent of the renormalization scheme (but depends on the coupling parameter $\xi$). It is a tensor covariantly constructed out of (covariant derivatives of) the Riemann tensor. 

Restricting to continuous self-similarity, \eqref{eq:MainResult} implies that
\beq
\label{eq:MainResultCSS}
 \langle T_{\mu \nu} \rangle( u, y ) = e^{2 u} \left( \overline{ \langle T_{\mu \nu} \rangle }(y) + \frac{u}{4\pi} \overline{V_{\mu \nu}}(y) \right),
\eeq
where $\overline{ \langle T_{\mu \nu} \rangle }(y)$ and $\overline{ V_{\mu \nu} }(y)$ are obtained by evaluation of $\langle T_{\mu \nu} \rangle$ and $V_{\mu \nu}$ at $u = 0$ (for discrete self-similarity, these two functions would be periodic in $u$). Essentially the same result on the stress tensor was obtained by Brady and Ottewill \cite{BradyOttewill} under the additional assumption of conformal coupling ($\xi = \frac{1}{6}$). Our result shows that conformal coupling (as well as continuous self-similarity) is not essential. As the scalar field relevant to self-similarity is typically not conformally (but minimally) coupled, our result should be a useful generalization.

In order to discuss the consequence of our result, we restrict to continuous self-similarity, i.e., consider \eqref{eq:MainResultCSS}. Following \cite{BradyOttewill}, one parameterizes metric perturbations in the form
\beq
 \tilde g_{\mu \nu}(u,y) = g_{\mu \nu}(u,y) + \delta g_{\mu \nu}(u,y) = e^{- 2 u} \left( g_{\mu \nu}(y) + e^{\omega_q u} \delta g_{\mu \nu}(y) \right),
\eeq
where $\tilde g_{\mu \nu}$ denotes the perturbed metric, $g_{\mu \nu}$ the unperturbed self-similar metric (again, $g_{\mu \nu}(y)$ is obtained from $g_{\mu \nu}(u, y)$ by evaluation at $u = 0$), and $\omega_q$ is the Lyapunov exponent characterizing the exponential growth of a perturbation of the self-similar metric induced by the quantum stress tensor. Plugging this into the left hand side of the semi-classical Einstein equation \eqref{eq:SCEE} and expanding to first order in $\delta g_{\mu \nu}$, one sees that one must have $\omega_q = 2$ in order to match the $u$ dependence on both sides of the equation. In fact, due to the supplementary linear growth in $u$ of the stress tensor, the above ansatz has to be slightly modified to also include a supplementary linear growth in $u$. We refer to \cite{BradyOttewill} for a detailed discussion and also of consequences of $\omega_q = 2$.

It is striking that for large $u$ the leading coefficient of $e^{2u}$ is state-independent and of geometric nature. This opens the possibility to incorporate quantum effects in self-similar collapse without actually computing vacuum expectation values. It should however be remarked that, when backreaction via the semi-classical Einstein equation is taken into account, the spacetime will in general cease to be self-similar, so that the assumption on which our analysis is built breaks down. Nevertheless, our result should be useful to study the regime of weak backreaction, where the deviations from self-similarity are yet small. In particular, it should provide a useful consistency check for approaches aiming at consistent solutions of the semi-classical Einstein equations (such as \cite{BercziEtAl2022}).

The universal scaling behaviour \eqref{eq:MainResult} (or its consequence \eqref{eq:MainResultCSS} in the continuously self-similar case) of the quantum stress tensor is reminiscent of a universality result for the quantum stress near the Cauchy horizon of Reissner-Nordstr\"om-deSitter spacetime: It was proven \cite{HollandsWaldZahn, HintzKlein} that in any state which is Hadamard (also across the cosmological horizon), one has
\beq
\label{eq:sCC_Universality}
 \EV{T_{VV}} = C V^{-2} + \cO(V^{-2 + 2 \beta}),
\eeq
where $V$ is the Kruskal coordinate which extends across the Cauchy horizon (and vanishes there), $C$ is state independent, and $\beta = \frac{\alpha}{\kappa_-}$ the ratio of the spectral gap of quasinormal modes and the surface gravity of the Cauchy horizon. One obvious difference to our result is that, contrary to the tensor $V_{\mu \nu}$ occurring in \eqref{eq:MainResult}, the coefficient $C$ is not determined by the local geometry, but by the scattering coefficients for mode scattering on the black hole exterior and interior region. Furthermore, while the result \eqref{eq:sCC_Universality} is rigorously proven for general Hadamard states, our result \eqref{eq:MainResult} holds as such only for states respecting self-similarity. In the Conclusion, we sketch arguments why the leading term proportional to $u e^{2 u}$ should be the same in a much broader class of Hadamard states, but we do not attempt to precisely define such a class or rigorously prove this statement. In any case, we think it is remarkable that both in the context of strong and weak cosmic censorship universal behaviour of quantum fields seems to be relevant.

The paper is organised as follows: In the next section, we prove the result \eqref{eq:MainResult} for the stress tensor in self-similar spacetimes. In Section~\ref{sec:SelfSimilarStates}, we present examples of states respecting self-similarity. One of these is generic enough to generalize to other self-similar spacetimes. We conclude in Section~\ref{sec:Conclusion} with a summary and a comparison of our results with those recently obtained in \cite{TomasevicWu} in a dimensional reduction approach.

\subsubsection*{Notation and conventions}

We use the ``mostly plus'' signature and the conventions of \cite{WaldGR} regarding curvature tensors. We work in units such that $G = c = \hbar = 1$.

\section{The renormalized stress tensor in self-similar spacetimes}

We recall the expression (here $G_{\mu \nu}$ is the Einstein tensor)
\beq
 T_{\mu \nu} = \nabla_{\mu} \phi \nabla_\nu \phi - \frac{1}{2} g_{\mu \nu} \nabla^\lambda \phi \nabla_\lambda \phi + \xi \left( g_{\mu \nu} \nabla^\lambda \nabla_\lambda \phi^2 - \nabla_\mu \nabla_\nu \phi^2 + G_{\mu \nu} \phi^2 \right)
\eeq
for the stress tensor of the massless scalar field. As it involves products of fields at coinciding points, the corresponding expression in the quantum theory requires renormalization. The latter should be performed in a local and covariant manner, a notion that is formulated axiomatically in \cite{HollandsWaldWick, HollandsWaldStress} (the axioms given there supersede Wald's axioms \cite{Wald77} for the stress tensor). Any renormalization scheme fulfilling the mentioned requirements can be shown to be of the Hadamard point-split form (see, for example, Section~2 of \cite{FrobZahn2019} for a detailed account), so that the renormalized expectation value of the stress tensor in a state with two-point function $w$ is given by 
\begin{multline}
\label{eq:HadamardPointSplit}
 \langle T_{\mu \nu}(x) \rangle = \lim_{x' \to x} \left( \nabla_\mu \nabla'_\nu - \frac{1}{2} g_{\mu \nu} \nabla^\lambda \nabla'_\lambda \right) \left( w(x; x') - h(x; x') \right) \\ + \xi \left( g_{\mu \nu} \nabla^\lambda \nabla_\lambda - \nabla_\mu \nabla_\nu + G_{\mu \nu}(x) \right) \lim_{x' \to x} \left( w(x; x') - h(x; x') \right) - C_{\mu \nu}(x).
\end{multline}
Here the primed derivatives act on the primed variable, $h(x; x')$ is the locally and covariantly constructed Hadamard parametrix, and the two-point function $w(x; x')$ is assumed to be of Hadamard form, i.e., such that $w - h$ is smooth near coinciding points. The tensor $C_{\mu \nu}$ is locally and covariantly constructed out of (covariant derivatives of) the curvature tensor and plays two roles: It ensures the conservation of the stress tensor expectation value, $\nabla^\mu \langle T_{\mu \nu} \rangle = 0$, and it incorporates finite renormalization ambiguities. In the massless case that we are considering, the latter consist of linear combinations of the two tensors
\begin{align}
 \delta T^{(1)}_{\mu \nu} & \defeq 4 \nabla_\mu \nabla_\nu R - 4 g_{\mu \nu} \nabla^\lambda \nabla_\lambda R - 4 R R_{\mu \nu} + g_{\mu \nu} R^2, \\
 \delta T^{(2)}_{\mu \nu} & \defeq 2 \nabla_\mu \nabla_\nu R - g_{\mu \nu} \nabla^\lambda \nabla_\lambda R - 2 \nabla^\lambda \nabla_\lambda R_{\mu \nu} + g_{\mu \nu} R_{\lambda \rho} R^{\lambda \rho} - 4 R_{\mu \lambda \nu \rho} R^{\lambda \rho},
\end{align}
which are the ``stress tensors'' obtained by variation with respect to $g^{\mu \nu}$ of the actions for the Lagrangians $R^2$ and $R_{\mu \nu} R^{\mu \nu}$. The precise form of $C_{\mu \nu}$ is irrelevant for our purposes, what matters is its behaviour under a global scale transformation $g_{\mu \nu}^{(\eta)} = \eta^2 g_{\mu \nu}$, which is given by $C_{\mu \nu}^{(\eta)} = \eta^{-2} C_{\mu \nu}$. It follows that in a self-similar spacetime, characterized by \eqref{eq:SelfSimilarity}, it fulfills
\beq
 C_{\mu \nu}(u + \alpha, y) = e^{2 \alpha} C_{\mu \nu}(u, y).
\eeq
Again, this may hold for $\alpha$ being an integer multiple of a period $\Delta$ (discrete self-similarity), or for arbitrary $\alpha$ (continuous self-similarity).

The Hadamard parametrix $h(x; x')$ entering the point-split prescription \eqref{eq:HadamardPointSplit} is locally (for $x'$ in a suitable neighborhood of $x$) of the form (omitting an $i \eps$ prescription irrelevant for our purposes)
\beq
\label{eq:Parametrix}
 h(x; x') = \frac{1}{8 \pi} \left( \frac{U(x; x')}{\sigma(x; x')} + V(x; x') \ln \frac{ \sigma(x; x')}{\Lambda^2} \right),
\eeq
where $\sigma(x; x')$ is Synge's world function (the signed squared geodesic distance of $x$ and $x'$ divided by 2, see also \cite{PoissonPoundVega}), $U(x; x')$ and $V(x; x')$ are smooth functions constructed locally and covariantly, and $\Lambda$ is an arbitrary but fixed scale. Under a global scale transformation $g_{\mu \nu}^{(\eta)} = \eta^2 g_{\mu \nu}$, the functions $\sigma$, $U$, and $V$ change as (the transformation behaviour of $U$ and $V$ follows straightforwardly from the transport equations defining them, \cf \cite{DecaniniFolacci08}, for example)
\begin{align}
 \sigma^{(\eta)} & = \eta^2 \sigma, &
 U^{(\eta)} & = U, &
 V^{(\eta)} & = \eta^{-2} V.
\end{align}
It follows that on a self-similar spacetime, the Hadamard parametrix fulfills
\beq
\label{eq:ParametrixScaling}
 h( u + \alpha, y; u' + \alpha, y') = e^{2 \alpha} \left( h(u, y; u', y') - \frac{\alpha}{4\pi} V(u, y; u', y') \right),
\eeq
with the inhomogeneous term due to the logarithm in \eqref{eq:Parametrix}.
Hence, assuming that the state respects self-similarity in the sense that \eqref{eq:2ptSelfSimilar} holds, we obtain
\begin{multline}
 \langle T_{\mu \nu}(u + \alpha, y) \rangle = e^{2 \alpha} \bigg( \langle T_{\mu \nu}(u, y) \rangle + \frac{\alpha}{4\pi} \bigg\{ [ \nabla_\mu \nabla'_\nu V](u, y) - \frac{1}{2} g_{\mu \nu} [\nabla^\lambda \nabla'_\lambda V](u,y) \\
 + \xi \left( g_{\mu \nu} \nabla^\lambda \nabla_\lambda - \nabla_\mu \nabla_\nu + G_{\mu \nu}(u,y) \right) [V](u, y) \bigg\} \bigg),
\end{multline}
where the square brackets denote the limit of coinciding points. We thus get \eqref{eq:MainResult} with
\beq
 V_{\mu \nu} = [ \nabla_\mu \nabla'_\nu V] - \frac{1}{2} g_{\mu \nu} [\nabla^\lambda \nabla'_\lambda V]+ \xi \left( g_{\mu \nu} \nabla^\lambda \nabla_\lambda - \nabla_\mu \nabla_\nu + G_{\mu \nu} \right) [V].
\eeq
Using results on the expansion of $V(x; x')$ near coinciding points \cite{DecaniniFolacci08}, one straightforwardly obtains
\beq
 V_{\mu \nu} = \left( \frac{1}{720} - \frac{(6\xi-1)^2}{288} \right) \delta T_{\mu \nu}^{(1)} - \frac{1}{240} \delta T_{\mu \nu}^{(2)}.
\eeq
As discussed in the Introduction, this coincides with the ambiguity of the stress tensor due to a change of the scale $\Lambda$ in the parametrix \eqref{eq:Parametrix}, which has been computed for example in \cite{Christensen76} (as the logarithmically divergent part of $T_{\mu \nu}$) or in \cite{HackThesis}, Theorem IV.2.1. For conformal coupling ($\xi = \frac{1}{6}$), $V_{\mu \nu}$ reduces to (a multiple of) the Bach tensor, so that we recover the results of \cite{BradyOttewill} for that case.

As the tensor $V_{\mu \nu}$ represents a renormalization ambiguity of the stress tensor, it is automatically conserved. However, $u V_{\mu \nu}(x)$ as it occurs in expression \eqref{eq:MainResultCSS}, is in general not conserved, due to the supplementary factor $u$. As the divergence of $V_{\mu \nu}$ vanishes, $\nabla^\mu( u V_{\mu\nu} ) = \nabla^\mu u V_{\mu \nu}$, which can be written as $e^{4 u} W_\nu(y)$ (due to the raised index in the derivative and $V_{\mu \nu}(x) = e^{2 u} \overline{V_{\mu\nu}}(y)$). As the full stress tensor is conserved, it follows that the first term on the right hand side of \eqref{eq:MainResultCSS}, $e^{2 u} \overline{\langle T_{\mu \nu} \rangle}(y)$, will also not be conserved, but its divergence equals $- \frac{1}{4 \pi} e^{4 u} W_\nu(y)$. This will also be discussed in examples below.

The analysis presented above concerns the physical spacetime dimension four. For general spacetime dimension $D$, in the condition \eqref{eq:2ptSelfSimilar} characterizing states respecting self-similarity, the scaling factor $e^{2 \alpha}$ has to be generalized to $e^{(D-2) \alpha}$, as this is the only scaling requirement that can consistently be imposed (as the antisymmetric part, given by the difference of retarded and advanced propagator, scales in this way). The Hadamard parametrix fulfills the analogous scaling, i.e., \eqref{eq:ParametrixScaling} with $e^{2 \alpha}$ replaced by $e^{(D-2) \alpha}$, but the function $V(x; x')$ is dimension dependent. In particular, it vanishes in odd spacetime dimension $D$. With the replacement of $e^{2 \alpha}$ by $e^{(D-2) \alpha}$ the general result \eqref{eq:MainResult} thus holds in general spacetime dimension $D$. However, the geometric quantity $V_{\mu \nu}(x)$ depends on the spacetime dimension, and in particular vanishes for odd $D$.

\section{States respecting self-similarity}
\label{sec:SelfSimilarStates}

We want to give three examples of states respecting self-similarity. The first one is the Minkowski vacuum state, restricted to a patch of Minkowski space which is self similar. This patch is given, in spherical coordinates, by the restriction $t < r$. With coordinates $(u, \tau)$ defined by
\begin{align}
 r & = \tau e^{-u}, &
 t & = ( \tau - 1) e^{-u},
\end{align}
the metric assumes the continuously self-similar form
\beq
 \ud s^2 = e^{- 2 u} \left( - (1 - 2 \tau ) \ud u^2 - 2 \ud u \ud \tau + \tau^2 \ud^2 \Omega \right),
\eeq
with $\ud^2 \Omega$ the metric on the unit sphere.
Now under $u \to u + \alpha$, both $r$ and $t$ are multiplied by the factor $e^{- \alpha}$. As the vacuum two-point function on Minkowski space is given by (we omit the $i \eps$ prescription, which is irrelevant for our argument)
\beq
 w(x; x') = - \frac{1}{4 \pi^2} \frac{1}{ (t - t')^2 - ( r^2 + {r'}^2 + 2 r r' \cos \theta ) },
\eeq
with $\theta$ the angle between $\mathbf{x}$ and $\mathbf{x}'$ (the spatial parts of $x$ and $x'$), this two-point function clearly has the desired self-similarity \eqref{eq:2ptSelfSimilar}. As the tensor $V_{\mu \nu}$ vanishes on Minkowski space, this is not a particularly interesting example, its main purpose being to show that the assumption of states respecting self-similarity is not overly restrictive.

Our second, more interesting, example concerns the minimally coupled scalar field on the Hayward spacetime \cite{HaywardSpacetime}. This is a solution to the Einstein-scalar field equation with minimal coupling. It is a special case of the family of Roberts spacetimes \cite{Roberts1989} (we refer to Section IV.E of \cite{GundlachHilditchMartin-Garcia} and Appendix C of \cite{TomasevicWu} for discussions of its relevance to critical collapse). It can be written in the form
\begin{align}
 \ud s^2 & = e^{-2 u} \left( - 2 \ud \tau^2 + 2 \ud u^2 + \ud^2 \Omega \right), & \Phi & = \tau,
\end{align}
with $\Phi$ the classical scalar field. Note that it is not conformal to Minkowski space, but to the two-sphere times two-dimensional Minkowski space. There is a curvature singularity at $u \to + \infty$.

Now consider fluctuations $\phi$ around $\Phi$ (with the spacetime fixed). The corresponding equation of motion is
\beq
\label{eq:eomHayward}
 \left( - \del_\tau^2 + e^{2 u} \del_u e^{-2 u} \del_u + 2 \Delta^{S_2} \right) \phi = 0,
\eeq
with $\Delta^{S_2}$ the Laplacian on the sphere. As the spacetime is static \wrt the Killing field $\del_\tau$, it is natural to use the ansatz ($Y_{\ell m}$ being spherical harmonics)
\beq
 \phi = e^{-i \omega \tau} Y_{\ell m}(\theta, \varphi) e^{u} R(u),
\eeq
for which the equation \eqref{eq:eomHayward} reduces to
\beq
 R''(u) - \left( 2 \ell (\ell + 1) + 1 - \omega^2 \right) R(u) = 0.
\eeq
This has the solution $R_k = e^{i k u}$, where $\omega$ and $k$ are related by
\beq
 \omega^2 = k^2 + 2 \ell (\ell + 1) + 1. 
\eeq

The relevant structure for the quantization of the field $\phi$ is the symplectic form of solutions to the field equation \eqref{eq:eomHayward}, which is given by
\beq
 \sigma( \phi_1, \phi_2 ) = \int \left( \phi_1 \del_\tau \phi_2 - \phi_2 \del_\tau \phi_1 \right) e^{-2 u} \ud u \ud \Omega,
\eeq
where the integration is over some constant $\tau$ slice.
Hence, the modes
\beq
 \phi_{k \ell m} = \frac{1}{\sqrt{4 \pi \omega}} e^{u} e^{i k u} e^{- i \omega \tau} Y_{\ell m}(\theta, \varphi)
\eeq
are symplectically normalized for real $k$, in the sense that
\beq
\label{eq:SymplecticNormalization}
 \sigma( \bar \phi_{k \ell m}, \phi_{k' \ell' m'} ) = - i \delta(k - k') \delta_{\ell \ell'} \delta_{m m'}.
\eeq
The set of modes is also complete. The corresponding Wightman two-point function is
\beq
\label{eq:2ptHayward}
 w(x; x') = \langle \phi(x) \phi(x') \rangle = \sum_{\ell, m} \int_{- \infty}^\infty \ud k  \frac{1}{4 \pi \omega} e^{- i \omega(\tau - \tau')} e^{i k (u - u')} e^{u + u'} Y_{\ell m}(\theta, \varphi) \bar Y_{\ell m}(\theta', \varphi').
\eeq
As $\del_\tau$ is a timelike Killing field and the modes used in this construction are of positive frequency with respect to $\del_\tau$, the state thus constructed represents the vacuum state (with respect to the time evolution given by $\del_\tau$). In particular, it is Hadamard \cite{SahlmannVerchPassivity} (i.e., with two-point function of Hadamard form), so the expectation value of the stress tensor can be defined by the Hadamard point-split procedure \eqref{eq:HadamardPointSplit}. Furthermore, the state respects self-similarity as \eqref{eq:2ptSelfSimilar} holds.

As the two-point function \eqref{eq:2ptHayward} is expressed as a mode integral involving very elementary functions, it is conceivable that the expectation value of the stress tensor in Hadamard point-splitting can be evaluated, for example by adapting the methods used in \cite{ArrecheaEtAl2024}. However, using the fact that the state respects self-similarity, we can use our general result \eqref{eq:MainResultCSS} to straightforwardly compute the leading contribution for large $u$, which is determined by $V_{\mu \nu}$. For minimal coupling $\xi = 0$, it is given by 
\begin{multline}
 V_{\mu \nu} = - \frac{1}{60} \nabla_\mu \nabla_\nu R + \frac{1}{80} g_{\mu \nu} \Box R + \frac{1}{120} \Box R_{\mu \nu} + \frac{1}{120} R R_{\mu \nu} - \frac{1}{480} g_{\mu \nu} R^2 \\ - \frac{1}{240} g_{\mu \nu} R_{\lambda \rho} R^{\lambda \rho} + \frac{1}{60} R_{\mu \lambda \nu \rho} R^{\lambda \rho}.
\end{multline}
For the Hayward spacetime, this can straightforwardly be evaluated to (here we are using the notation employed in \eqref{eq:MainResultCSS})
\begin{align}
\label{eq:V_Hayward}
 \overline{V_{\tau \tau}} & = \frac{1}{80}, & \overline{V_{u u}} & = \frac{17}{240}, & \overline{V_{\theta \theta}} & = - \frac{7}{480}, & \overline{V_{\phi \phi}} & = \overline{V_{\theta \theta}} \sin^2 \theta,
\end{align}
with all other components vanishing. We note that $V_{\mu \nu}$ is traceless. This is consistent with the fact that, as discussed in the Introduction, $V_{\mu \nu}$ represents the renormalization ambiguity associated to global scale transformations. But we know that the renormalization ambiguity in a massless theory has a trace proportional to $\Box R$ (as both $\delta T_{\mu \nu}^{(1)}$ and $\delta T_{\mu \nu}^{(2)}$ have a trace proportional to this expression). As $R = - e^{2 u}$, this is easily seen to vanish. 

As mentioned above, the divergence $\nabla^{\mu} (u V_{\mu \nu}) = e^{4 u} W_\nu$ does not vanish in general. Concretely, one computes $W_u = \frac{17}{480}$, with all other components vanishing. By spherical symmetry and the fact that we are using the ground state \wrt $\del_\tau$, it follows that $\overline{ \langle T_{\mu \nu } \rangle }$ is of the form
\begin{align}
 \overline{\EV{T_{\tau \tau}}} & = a, & \overline{\EV{T_{u u}}} & = b, & \overline{\EV{T_{\theta \theta}}} & = c, & \overline{\EV{T_{\phi \phi}}} & = \overline{\EV{T_{\theta \theta}}} \sin^2 \theta,
\end{align}
with constants $a$, $b$, $c$ and all other components vanishing. Computing the divergence of $e^{2 u} \overline{\EV{T_{\mu \nu}}}$ and using that the divergence of the total stress tensor vanishes, we obtain the relation
\beq
 a - b - 4 c = \frac{1}{4 \pi} \frac{17}{240}
\eeq
between the parameters $a$, $b$, $c$. This leaves us with two parameters, which corresponds to the number of free parameters due to the renormalization ambiguity (the two tensors $\delta T^{(1)}_{\mu \nu}$ and $\delta T^{(2)}_{\mu \nu}$ are linearly independent on the Hayward spacetime). Hence, we have determined the full stress tensor expectation value, including its renormalization ambiguity.

We now turn to our third example, the critical Roberts spacetime \cite{Roberts1989} (\cf Section IV.E of \cite{GundlachHilditchMartin-Garcia} and Section 5.1 of \cite{TomasevicWu} for discussions of its relevance to critical collapse). In coordinates which make the self-similarity explicit, its metric is \cite{Frolov1997}
\beq
\label{eq:CriticalRoberts}
 \ud s^2 = e^{-2 (u - r)} \left( 2 (1 - e^{-2 r}) \ud u^2 - 4 \ud u \ud r + \ud^2 \Omega \right).
\eeq
Here $u \in \R$, but $r > 0$ (at $r = 0$, one can continuously match a patch of Minkowski space). It follows that the surfaces of constant $u$ are null hypersurfaces, while those of constant $r$ are spacelike and in fact Cauchy surfaces. The spacetime is not asymptotically flat. In the large $r$ limit (so that $e^{-2r} \ll 1$), one recovers the Hayward spacetime, which becomes explicit through the coordinate change
\begin{align}
\label{eq:CoordinateTrafoRobertsHayward}
 \tilde \tau & = r, & \tilde u & = u - r.
\end{align}
In particular, asymptotically (large $r$), $\tilde u = u - r$ is also a suitable self-similarity coordinate.

The Klein-Gordon operator corresponding to the metric \eqref{eq:CriticalRoberts} is
\beq
  - \frac{1}{2} e^{4(u-r)} \left( \del_r (1- e^{-2r} )e^{-2(u-r)} \del_r + \del_u e^{-2(u-r)} \del_r + \del_r e^{-2(u-r)} \del_u \right)  + e^{2 (u - r)} \Delta^{S_2}.
\eeq
A suitable mode ansatz is
\beq
\label{eq:ModeAnsatzRoberts}
 \phi = e^{u-r} e^{- i k u} R(r) Y_{\ell m}(\theta, \phi),
\eeq
which leads to the equation
\beq
 - ( 1 - e^{- 2 r} ) R''(r) + 2 (i k - e^{- 2 r} ) R'(r) - ((1 + 2 \ell (\ell + 1)) - e^{- 2 r}) R(r) = 0.
\eeq
This can be solved in terms of the hypergeometric function, but the concrete form of these solutions is irrelevant for our purposes. More relevant for us is the behaviour at large $r$, for which $R(r)$ asymptotically solves
\beq
 - R''(r) + 2 i k R'(r) - (1 + 2 \ell (\ell + 1)) R(r) = 0,
\eeq
which is solved by a linear combination of $e^{ i ( k \mp \omega ) r}$, where
\beq
 \omega = \sqrt{ k^2 + 2 \ell (\ell +1) + 1}.
\eeq

We must consider those modes which are symplectically normalizable. The symplectic form on the $r = \const$ hypersurfaces is given by
\beq
 \sigma(\phi_1, \phi_2) = \int_{-\infty}^\infty \ud u \int \ud \Omega \ e^{2 (r - u)} \left\{ \phi_1 \left( \del_u + (1 - e^{-2 r}) \del_r \right) \phi_2 - \phi_1 \leftrightarrow \phi_2 \right\}.
\eeq
For the modes \eqref{eq:ModeAnsatzRoberts}, symplectic normalization in the sense of \eqref{eq:SymplecticNormalization} is obviously only possible for real $k$, as otherwise the integral diverges either for $u \to \infty$ or $u \to - \infty$. Considering the asymptotic behaviour $R(r) \simeq e^{ i ( k \mp \omega ) r}$ for large $r$ and the fact that $\del_r$ is a future pointing null vector, this suggests to choose the solution with the upper sign in order to have positive energy modes in the asymptotic region of large $r$. As the symplectic form is conserved (independent of $r$), we can evaluate it at any $r$, in particular in the asymptotic case $r \to \infty$. Normalizing $R_{k \ell}$ such that $R_{k \ell}(r) \simeq e^{i (k - \omega) r}$ in that regime, we thus get the symplectically normalized modes as
\beq
 \phi_{k \ell m} = \frac{1}{\sqrt{4 \pi \omega}} e^{u-r} e^{- i k u} R_{k \ell}(r) Y_{\ell m}(\theta, \phi).
\eeq
The corresponding Wightman two-point function
\beq
\label{eq:ModeSumGeneral}
 w(x; x') = \int_{- \infty}^\infty \ud k \sum_{\ell, m} \phi_{k \ell m}(x) \overline{ \phi_{k \ell m} (x') }
\eeq
then obviously fulfills the condition \eqref{eq:2ptSelfSimilar} for a state respecting self-similarity. Using the coordinate transformation \eqref{eq:CoordinateTrafoRobertsHayward} it is also obvious that the state thus constructed asymptotically (for large $r$) approaches the vacuum state on the Hayward spacetime. In particular, the constructed state is Hadamard (see \cite{GerardWrochnaAsymptoticallyStatic} for a proof of the Hadamard property of in/out states on asymptotically static spacetimes, where only a power law convergence is required, while here we even have exponential convergence).

For the tensor $V_{\mu \nu}$ we obtain
\begin{align}
\label{eq:V_munu_Roberts}
 \overline{V_{u u}} & = e^{- 2 r} \frac{1}{240} (1 - e^{-2r} ) ( 17 - 18 e^{-2r} + 45 e^{-4 r} ), &
 \overline{V_{r r}} & = e^{- 2 r} \frac{1}{60} (5 - 17 e^{-2 r} ), \\
 \overline{V_{u r}} & = - e^{- 2 r} \frac{1}{240} ( 17 - 58 e^{- 2 r} + 45 e^{- 4 r} ), &
 \overline{V_{\theta \theta}} & = - e^{- 2 r} \frac{1}{480} ( 7 + 26 e^{- 2 r} - 69 e^{- 4 r} ), \nn
\end{align}
and $\overline{V_{\theta \theta}} = \sin^2 \theta \overline{V_{\theta \theta}}$, with all other components vanishing. 
From the fact that our state asymptotes to the state that we constructed on the Hayward spacetime, we know that we must have, in the coordinates introduced in \eqref{eq:CoordinateTrafoRobertsHayward},
\begin{align}
 \EV{ T_{\tilde \tau \tilde \tau} } & = e^{2 \tilde u} \left( a + \frac{1}{4 \pi} \frac{1}{80} \tilde u + \cO(e^{-2 r}) \right), \\
 \EV{ T_{\tilde u \tilde u} } & = e^{2 \tilde u} \left( b + \frac{1}{4 \pi} \frac{17}{240} \tilde u + \cO(e^{-2 r}) \right), \\
 \EV{ T_{\theta \theta} } & = e^{2 \tilde u} \left( \frac{a-b}{4} - \frac{1}{4 \pi} \frac{17 + 14 \tilde u}{960} + \cO(e^{-2 r}) \right),
\end{align}
with $a$, $b$ parameterizing the renormalization ambiguity. Converting the back into $(u, r)$ coordinates, this is consistent with \eqref{eq:V_munu_Roberts}. We also see that we have terms of the form $r e^{2 (u - r)}$ in the expectation value. In fact, the necessity of such terms can also be inferred from the required cancellation of the divergence of $u e^{2 u} \overline{V_{\mu \nu}}$.

The above construction corresponds to an ``out''-state, defined by data at large $r$. One could also define an ``in''-state by choosing positive frequency data at $r = 0$. This state respects self-similarity, too. Which of these states is considered more relevant will depend on the physical situation one aims to describe. In any case, the leading geometric term in the stress tensor expectation value is the same in any state respecting self-similarity.

The construction we presented for a state respecting self-similarity on the critical Roberts spacetime can be straightforwardly generalized to other continuously self-similar spherically symmetric spacetimes. When using coordinates $(u, r, \theta, \phi)$ where $u$ is the self-similarity time such that \eqref{eq:SelfSimilarity} holds and $r$ is such that the constant $r$ slices are Cauchy surfaces, then the mode ansatz
\beq
 \phi = e^u e^{- i k u} R(r) Y_{\ell m}(\theta, \phi)
\eeq
gives rise to a second order differential equation for $R(r)$ (the supplementary factor $e^{-r}$ in \eqref{eq:ModeAnsatzRoberts} was introduced for the concrete case of the critical Roberts spacetime to simplify the equation for $R(r)$). By the assumed self-similar form of the metric, the symplectic form defined by integration over constant $r$ slices, involves the integration measure $e^{- 2 u} \ud u$. Hence, the above modes will be symplectically normalizable only for real $k$. By taking all real $k$ and all $\ell, m$, one thus obtains a complete set of mode solutions. It remains to select ``positive frequency'' ones, for example by imposing initial or final data for $R(r)$. The resulting state defined by a mode sum as in \eqref{eq:ModeSumGeneral} then respects self-similarity.

\section{Conclusion}
\label{sec:Conclusion}

We showed that, in states respecting self-similarity, the quantum stress tensor in self-similar spacetimes can be characterized by a Lyapunov exponent $\omega_q = 2$. Moreover, the leading coefficient of the exponential growth is, for large $u$, determined only by local geometric data. Our result generalizes that of \cite{BradyOttewill} to non-conformal (but massless) fields. Moreover, we showed existence of states respecting self-similarity in three examples, (a patch of) Minkowski space, the Hayward spacetime, and the critical Roberts spacetime. We also indicated how this generalizes to other self-similar spacetimes.

We already motivated the assumption of a state respecting self-similarity, but nevertheless we would like to briefly discuss states not fulfilling that assumption. 
The difference of any two two-point functions of Hadamard form is a smooth symmetric bi-solution to the Klein-Gordon equation (a solution in both arguments). Assuming that there is a Hadamard state respecting self-similarity (and a general construction of such was sketched above), we can thus write
\begin{multline}
\label{eq:StateSubtraction}
 \EV{T_{\mu \nu}(x)} = \EV{T_{\mu \nu}(x)}_{\mathrm{s.s.}} +  \lim_{x' \to x} \left( \nabla_\mu \nabla'_\nu - \frac{1}{2} g_{\mu \nu} \nabla^\lambda \nabla'_\lambda \right) W(x;x') \\ + \xi \left( g_{\mu \nu} \nabla^\lambda \nabla_\lambda - \nabla_\mu \nabla_\nu + G_{\mu \nu}(x) \right) W(x;x),
\end{multline}
where $ \EV{ \cdot }_{\mathrm{s.s.}}$ denotes the expectation value in the self-similar state and $W(x;x') \defeq w(x;x') - w_{\mathrm{s.s.}}(x;x')$ is the difference of the two-point functions of the actually considered and the reference (respecting self-similarity) state. At first sight, it may thus seem as if one could easily get rid of the $u e^{2 u}$ term in stress tensor expectation value, simply by choosing $W(x;x')$ appropriately. For example, one may fix $W$, $\del_r W$, $\del_{r'} W$, and $\del_r \del_{r'} W$ on some $r = r' = \const$ slice (in coordinates as used in the critical Roberts case) to coincide with $V$ (and its corresponding derivatives), multiplied by $- \frac{u + u'}{8 \pi}$. Then the term $\frac{u}{4 \pi} \overline{V_{\mu \nu}}$ in the stress tensor expectation value is cancelled (albeit only at the given $r$). However, while such a construction is compatible with $W(x; x')$ being a smooth symmetric bi-solution, it obviously leads to a violation of the positivity constraint (by large shifts in $u$ in either positive or negative direction). It thus seems highly implausible that there are reasonable states in which the $u e^{2 u}$ term in the stress tensor expectation value is cancelled or modified, mainly due to the positivity restriction.

Let us compare our results on the scalar quantum field on the Hayward and the critical Roberts spacetime with those recently obtained in \cite{TomasevicWu}. There, a dimensional reduction technique is used, i.e., one first uses the assumption of spherical symmetry to reduce the classical theory to a two-dimensional one, involving a dilaton. In the two-dimensional theory, one has a classically traceless stress tensor, but a trace anomaly in the quantum theory. This leads to a one-loop non-local effective action. Due to the non-locality, boundary conditions are necessary, which amount to a state-dependence. Also in this framework, a Lyapunov exponent $\omega_q = 2$ is found (which, as argued in the Introduction based on the fixed scaling of the commutator and the positivity constraint, is a generic prediction of quantum field theory on curved spacetimes). The results of \cite{TomasevicWu} on the stress tensor differ from ours in the coefficient of $e^{2 u}$, as there is no geometric component growing linearly in $u$. As it is not claimed in \cite{TomasevicWu} that the considered state respects self-similarity, this is not in direct contradiction to our general statement about such states. However, the stress tensor computed in \cite{TomasevicWu} depends on $u$ only through $e^{2 u}$ and in view of the above discussion it seems implausible that there is a Hadamard state giving rise to such a stress tensor expectation value. From the perspective of local and covariant renormalization taken here, it seems as if the scaling effects of renormalization, which are responsible for the $u e^{2 u}$ term, are not properly accounted for in the dimensional reduction framework used in \cite{TomasevicWu}.

As for future perspectives, it would certainly be interesting to apply our general result for the study of backreaction. Due to its linear growth, the geometric term is generically the dominant one at large $u$, so that in a first approximation the state-dependent part could be ignored. 

Our construction of states respecting self-similarity on the critical Roberts spacetime uses Cauchy surfaces extending from $-u$ to $u$ at a fixed $r$. A construction of states using (asymptotic) characteristic data (for example at $r = 0$ and $u = - \infty$), similar to the construction of states in black hole spacetimes, might be perceived more natural, and thus also a worthwhile endeavour. More generally, exactly self-similar spacetimes are typically not asymptotically flat, so they should be cut off and glued into an asymptotically flat spacetime \cite{OriPiran1987}. Considering such constructions could also provide guidance about suitable constructions of states (via initial or characteristic data).

\subsubsection*{Acknowledgements}

I would like to thank the authors of \cite{TomasevicWu} for illuminating discussions on their work.


\end{document}